\documentclass[11pt,paper,a4paper]{JHEP3} 

\JHEPspecialurl{http://jhep.sissa.it/JOURNAL/JHEP3.tar.gz}

\usepackage{epsfig,multicol,bbm}
\usepackage{amssymb}
\usepackage{amsmath}
\usepackage{cite}

\newcommand\fverb{\setbox\pippobox=\hbox\bgroup\verb}
\newcommand\fverbdo{\egroup\medskip\noindent%
                        \fbox{\unhbox\pippobox}\ }
\newcommand\fverbit{\egroup\item[\fbox{\unhbox\pippobox}]}
\newbox\pippobox
\newcommand{\beq}{\begin{equation}}
\newcommand{\eeq}{\end{equation}}
\newcommand{\bea}{\begin{eqnarray}}
\newcommand{\eea}{\end{eqnarray}}
\newcommand{\bem}{\begin{multline}}
\newcommand{\eem}{\end{multline}}
\newcommand{\beg}{\begin{gather}}
\newcommand{\eeg}{\end{gather}}

\title{Reaction-diffusion processes in zero transverse dimensions as toy models for high-energy QCD}

\author{N\'estor Armesto$^1$, Sergey Bondarenko$^1$, Jos\'e Guilherme Milhano$^2$ and Paloma Quiroga$^1$

\vspace{0.1in}

{\it
 $^1$ Departamento de F\'{\i}sica de Part\'{\i}culas and
IGFAE,
Universidade de Santiago de Compostela,
 15782 Santiago de Compostela, Spain

$^2$ CENTRA, Instituto Superior T\'ecnico (IST),
Av. Rovisco Pais, P-1049-001 Lisboa, Portugal

\vskip 0.1in

E-mail addresses: {\tt nestor@fpaxp1.usc.es, sergey@fpaxp1.usc.es, gui@fisica.ist.utl.pt, pquiroga@fpaxp1.usc.es}
}
}

\abstract{We examine numerically different zero-dimensional reaction-diffusion processes as candidate toy models for high-energy QCD evolution. Of the models examined -- Reggeon Field Theory, Directed Percolation and Reversible Processes -- only the latter shows the behaviour commonly expected, namely an increase of the scattering amplitude with increasing rapidity. Further, we find that increasing recombination terms, quantum loops and the heuristic inclusion of a running of the couplings, generically slow down the evolution.}


\begin{document} 



\section{Introduction}
\label{intro}

The seminal work of Munier and Peschanski \cite{Munier:2003vc} established a solid link between high energy QCD evolution and the physics of statistical mechanics systems.
In that and ensuing works \cite{Munier:2003sj,Munier:2004xu}, the connection between the Balitsky-Kovchegov (BK) equation \cite{Balitsky:1995ub,Kovchegov:1999yj} and the Fisher-Kolmogorov-Petrovsky-Piscounov (FKPP) equation \cite{fisher,kpp}, used to describe the time evolution of certain quantities in some statistical mechanics systems, was exploited to give a general and elegant derivation of the energy dependence of the saturation scale and of geometric scaling \cite{Iancu:2002tr,Mueller:2002zm}.

The BK equation is the mean field limit of the more general set of evolution equations  -- the B-JIMWLK\footnote{Balitsky--Jalilian-Marian--Iancu--McLerran--Weigert--Leonidov--Kovner.} equations \cite{Balitsky:1995ub,McLerran:1993ni,McLerran:1993ka,McLerran:1994vd,Jalilian-Marian:1997dw,Kovner:2000pt,Iancu:2000hn,Ferreiro:2001qy} -- describing the evolution of QCD scattering amplitudes at high energy, i.e. the dynamics of the Colour Glass Condensate (CGC).
Most studies of high energy QCD evolution have been performed using the BK equation.
The reasons behind such preference are clear. Not only is the BK equation considerably more tractable than the B-JIMWLK set, but also the  differences between the full B-JIMWLK evolution scheme and its truncated BK version are, perhaps surprisingly, small. The detailed numerical evaluation of these differences was performed in \cite{Rummukainen:2003ns} by recasting the full evolution as a random walk in some functional space, once again highlighting the possible deep connection between QCD evolution and statistical mechanics.  

The B-JIMWLK evolution equations were derived under the explicit assumption of the scattering of a dilute projectile on a dense target and should, therefore, be applicable only for that physical situation.
Such a limitation renders the CGC framework -- either in its simplified BK version or for the more complete B-JIMWLK set -- insufficient to address the experimentally important collisions of two nuclei. From a more theoretical perspective, this incompleteness of B-JIMWLK has been shown to eventually lead to the violation of unitary constraints \cite{Iancu:2003uh,Mueller:2004se}.

The inclusion of the missing ingredients -- variously referred  to as pomeron loops, fluctuations or wave function saturation effects -- in an extended B-JIMWLK scheme has been attempted by several groups. A rather stringent constraint on the form of the complete evolution is imposed by what has been called the Dense-dilute Duality (DdD) \cite{Kovner:2005en,Kovner:2005uw}. This result, which establishes the equivalence of the dilute projectile, dense target case addressed by B-JIMWLK and its mirror image of a dense projectile and a dilute target, implies that the complete evolution kernel should necessarily be self-dual under the DdD.

Of particular relevance to the subject matter of this paper are those attempts \cite{Iancu:2004es,Iancu:2004iy,Janik:2004ve,Mueller:2005ut,Levin:2005au,Iancu:2005nj,Iancu:2005dx,Marquet:2005ak,Armesto:2006ee,Blaizot:2006wp,Munier:2006um,Iancu:2006jw,Bondarenko:2006rh,Bondarenko:2007kg} to include the missing effects which resort to possible  generalizations  of the correspondence of high energy QCD  and statistical mechanics beyond the proven instance of \cite{Munier:2003vc}.
It has been argued \cite{Iancu:2004es} that the incompleteness of B-JIMWLK can be related to the breakdown of the FKPP description in systems involving a small number of objects. Thus, a powerful tool for the generalization of B-JIMWLK could be provided by insights from statistical mechanics systems where the discreteness effects 
that violate the FKPP description can be accounted for via the inclusion of stochastic effects.

In the context of statistical mechanics systems, the failing FKPP description is remedied by the inclusion of a stochastic (noise) term accounting for the effects of number fluctuations which become increasingly important with decreasing number of intervening objects. 
Different choices for the stochastic term to be added to the FKPP equation will, in general, lead to different dynamical behaviours. 
The choice of the noise term is thus a crucial step in obtaining a suitable description of high energy QCD evolution in the framework of stochastic processes.

In general, there are two alternative strategies that can be followed to determine the form of the noise term. One can attempt to extend the known B-JIMWLK evolution equations by including the missing effects and then interpret the resulting equations in the stochastic language. This approach was followed in \cite{Mueller:2005ut,Iancu:2005dx} where the obtained noise terms were, unfortunately, far too complex to allow any computation -- analytical or numerical -- to be carried out.
Instead, one can choose to make an educated guess as to what the noise should be. Ultimately, for any such choice, the behaviour of the resulting Langevin equation will need to be compared with our general expectations for high energy evolution.
In this paper we follow this latter strategy.

Our analysis is performed in the general framework of reaction-diffusion processes, in particular in the Hamiltonian language of \cite{elka}, which is equivalent to the formulation in terms of Langevin equations.
A Langevin equation in which the  stochastic term is such that the functional form of the two-point noise correlation function
 replicates that of the non-diffusive part of the FKPP equation is usually referred to as the  stochastic FKPP (sFKPP) equation and describes the dynamics of a reaction-diffusion process belonging \cite{dms} to the universality class of the reversible processes.
If the noise is chosen to have a two-point correlator simply proportional to the quantity obeying the FKPP equation, the resulting Langevin equation belongs to a different universality class, that of directed percolation.
Although other choices for the form of the stochastic term are certainly possible, we concentrate on these two since they have been previously considered as possible descriptions of high energy QCD in the framework of stochastic processes.

We explore numerically the universality classes of reaction-diffusion processes with zero transverse dimensions.
The Hamiltonian approach of \cite{elka} allows us to address both reversible processes and directed percolation universality classes within a single formulation. Further, the same formulation is well suited to enable a discussion of Reggeon Field Theory -- both in its original form and with the extra quartic vertex recently considered in the context of high energy QCD evolution \cite{Bondarenko:2006rh,Braun:2006gy} -- which, although related to directed percolation \cite{cardysugar}, does not correspond to a well defined reaction-diffusion process. Finally, it also allows for a rather straightforward heuristic introduction of `running coupling effects'.
Zero-dimensional toy models are  thus  used as a test bed for the far more involved realistic case with two transverse dimensions.
By comparing the dynamical behaviour of the reaction-diffusion models with general expectations for high energy QCD evolution -- geometrical scaling in BK, increase of the scattering amplitude with increasing rapidity, slowdown of evolution once couplings are allowed to run, diffusive scaling once stochastic effects are included  -- we narrow down the number of possible candidate models. The central result of this paper is that, at least in relation to toy models with zero transverse dimensions, the only universality class of reaction-diffusion processes compatible with high energy QCD evolution is that of reversible processes.  

In Sec. \ref{formalism} we introduce the formalism of  \cite{elka}, the toy models to be  studied, an heuristic implementation of running of the couplings, and the classical solution of the problem. Our results are presented in Sec. \ref{results} and a concluding discussion in Sec. \ref{conclusions}.

\section{Formalism}
\label{formalism}
In this Section we start by presenting the basic formalism that will be used to discuss the different classes of reaction-diffusion models. Then we motivate a choice of parameters in the models, discuss an heuristic introduction of the running of the couplings, and the classical solutions to the problem.

\subsection{Basic setup}
\label{setup}

We use the formalism developed in \cite{elka}. In this reference, the diffusive term in a stochastic equation plays the role of a kinetic term in a Hamiltonian formulation, while the details of the process are encoded in a reaction Hamiltonian which corresponds to a potential.
Following the notation of \cite{elka}, the reaction Hamiltonian for an elementary reaction 
$kA \overset{\lambda}{\rightarrow} mA$, where  $\lambda$ is the strength of the reaction, reads
\begin{equation}
H_R(\bar p, \bar q) =\frac{\lambda}{k!}(\bar p^m -\bar p^k) \bar q^k\, , 
\label{eq:2-0}
\end{equation}
with $[\bar q,\bar p]=1$.

Instead of discussing arbitrary cases, we focus in this paper in Hamiltonians with two powers of $\bar{q}$ at most. While higher vertices might be required by an eventual `true' theory of high energy QCD, we restrict to this form which contains the vertices most often discussed in the literature. 
To pass to a more usual notation, we make the changes $\bar p \to -q$, $\bar q \to  p$, $H_R \to  -H$, arriving at
\begin{equation}
H(p,q)=\alpha_1pq-\alpha_2qp^2-\alpha_3q^2p+\alpha_4q^2p^2\, .
\label{eq:2-1}
\end{equation}
This expression corresponds to the Hamiltonian which rules, for reactions involving vertices up to fourth order, the evolution in rapidity $y$ (the logarithm of the energy) of an auxiliary function $F(y,q)$ \cite{Braun:2006gy},
\begin{equation}
\frac{\partial F(y,q)}{\partial y} = -H(p,q) F(y,q)\, .
\label{eq:2-2}
\end{equation}
The evolution starts from some functional form for the coupling to the projectile at $y=0$, e.g. an eikonal coupling
\begin{equation}
F(y=0,q)=1-\exp{(-g_i q)}\, ,
\label{eq:2-3}
\end{equation}
where  $g_i$ is the coupling parameter with the projectile. The relation with the transition amplitude $A_{fi}(y)$ is given through
\begin{equation}
iA_{fi}(y)=F(y,q=g_f)\, ,
\label{eq:2-4}
\end{equation}
where $g_f$ is the coupling parameter with the target.

In (\ref{eq:2-1}) the variable $p$ plays the role of a conjugate variable to $q$, $p=-\partial/\partial q\equiv -\partial_q$. It is, in the formalism of \cite{elka}, the auxiliary degree of freedom required to study the non-equilibrium problem (for other approaches see \cite{doi,peliti,janssen,msr}). Further, $q(p)$ can be interpreted as the creation (annihilation) operator of the objects exchanged in the scattering. Within this interpretation, the property of the Hamiltonian (\ref{eq:2-1}),
\begin{equation}
H(p,\partial_p)=H_{y\to -y}\left(\frac{\alpha_3}{\alpha_2}q,-\frac{\alpha_2}{\alpha_3}\partial_q  \right)\, ,
\label{eq:2-5}
\end{equation}
amounts to a  symmetry of the scattering process from the projectile and target points of view. This symmetry is strongly suggestive of the Dense-dilute Duality which has been postulated \cite{Kovner:2005en} to be an essential property of high energy QCD evolution Hamiltonians.

In \cite{elka} the phase portrait determined by the zero energy trajectories, $H_R(\bar p,\bar q)=0$, has been used to classify reaction-diffusion processes into universality classes. We will focus in the following cases:

\begin{enumerate}

\item Reggeon Field Theory (RFT): $\alpha_4=0$, $\alpha_2=\alpha_3$. It contains no quartic vertex and  the splitting and recombination strengths are equal. It has no correspondence to a reaction-diffusion process, and it displays  the known phenomenon that the amplitude vanishes with increasing rapidity which is usually interpreted as tunneling \cite{Alessandrini:1975ak,Ciafaloni:1977xv,Ciafaloni:1978nb}. The effect of the quartic vertex has been examined recently \cite{Bondarenko:2006rh} due to its possible connection to high energy QCD evolution. The zero energy lines determine a triangular phase portrait given by
\begin{equation}
\bar p=0,\ \ \bar q=0, \ \  \bar q= \frac{\alpha_1}{\alpha_2}+\bar p\, .
\label{eq:2-6}
\end{equation}

\item Directed Percolation (DP): it is the case for the stochastic process with allowed reactions $1\overset{\lambda}{\rightarrow} 0$, $1\overset{\mu}{\rightarrow} 2$ and $2\overset{2\sigma}{\rightarrow} 1$, leading to $\alpha_1=\mu-\lambda$, $\alpha_3=\mu$ and $\alpha_2=\alpha_4=\sigma$. It contains a quartic vertex with the same strength as the recombination vertex, and its phase portrait is determined by
\begin{equation}
\bar p=0,\ \ \bar q=0, \ \  \bar q= \frac{\alpha_1+\alpha_3\bar p}{\alpha_2(1+\bar p)}\,. 
\label{eq:2-7}
\end{equation}
Renormalization group arguments lead to the conclusion that the quartic vertex is irrelevant at dimensions greater than 4 (see e.g.\cite{Tauber:2005ax}). On the other hand, the third trajectory in (\ref{eq:2-7}) goes into a straight line when expanded around $\bar p=0$. In both cases one returns to RFT, a fact that we will use below to fix the parameters for our numerical study.

\item Reversible processes (RP): it is the case for the stochastic process with reactions $1\overset{\mu}{\rightarrow} 2$ and $2 \overset{2\sigma}{\rightarrow} 1$, leading to $\alpha_1=\alpha_3=\mu$ and $\alpha_2=\alpha_4=\sigma$. It contains a quartic vertex with the same strength as the recombination vertex and its phase portrait is determined by
\begin{equation}
\bar p=0,\ \ \bar q=0, \ \ \bar p = -1, \ \  \bar q= \frac{\alpha_1}{\alpha_2}\,. 
\label{eq:2-8}
\end{equation}
This phase portrait cannot be deformed to a triangle and it does, therefore, belong  to a different universality class than that of DP. 
The Langevin equation corresponding to this reaction-difussion problem is the stochastic Fisher-Kolmogorov-Petrovsky-Pis\-cou\-nov (see recent analyses in \cite{jack,alhammal}), which has been proposed to play a role in high energy QCD evolution beyond JIMWLK \cite{Iancu:2004iy,Iancu:2005nj,Mueller:2005ut,Bondarenko:2007kg}.

\end{enumerate}

\subsection{Choice of parameters}
\label{param}

Instead of scanning a large region of the parameter space, we focus on the effect of the quartic vertex and of the difference between different reaction-diffusion processes. Thus, for RFT we choose values $\alpha_1=1$ (this parameter can always be absorbed in a redefinition of rapidity) and $\alpha_2=\alpha_3=0.5$. We have checked that with this set of parameters the conclusions we extract by numerically solving the evolution equation (\ref{eq:2-2}) in a restricted rapidity window $0<y<5$ hold for higher rapidities and are not a `subasymptotic' effect, a situation which may happen for smaller values of $\alpha_2=\alpha_3$ \cite{Braun:2006gy}.
Besides, we will generically study the introduction of a quartic vertex in RFT as done in \cite{Bondarenko:2006rh} though this situation, as RFT itself in zero dimensions, does not correspond to any reaction-diffusion process.

For DP we adopt the following strategy: we approximate the phase portrait (\ref{eq:2-7}) by a triangle, which brings DP into RFT with a given quartic vertex, and identify the parameters with those in the phase portrait of RFT (\ref{eq:2-6}). This leads to the values $\alpha_1=1$, $\alpha_2=\alpha_4=0.5$ and $\alpha_3=1.5$.

Finally, for RP we take $\alpha_1=\alpha_3=1$ and vary the remaining parameter in the region $0.1<\alpha_2=\alpha_4<0.9$.

\subsection{Running of the couplings}
\label{running}

All couplings in these reaction-difussion processes are, at this level, fixed. Recently it has been proposed in the framework of a one-dimensional model, that effects of the running of the couplings may shift the contribution of effects of loops to higher rapidities \cite{Dumitru:2007ew}. To implement the running of the coupling in our approach we adopt an heuristic procedure. We interpret $1/q$ as some `momentum' scale which determines a common logarithmic running for the four couplings in our Hamiltonian\footnote{This choice of scale can be argued as follows: we identify the amplitude with that for the scattering of a dipole of size $\propto q$, which is known to increase with increasing dipole size. Previous experience with running coupling effects in the BK equation \cite{Albacete:2004gw} give us the hope that the concrete choice of scale will not alter the qualitative conclusions of this study.}:
\begin{equation}
\alpha_i(q)=\alpha_i \,\frac{\ln{(Q/q)}}{\ln{(q_0/q)}},\ \ i=1,\dots ,4\, ,
\label{eq:2-9}
\end{equation}
for $q<q_0$ and frozen to the values $\alpha_i$ discussed in the previous Subsection for $q\ge q_0$. $Q=10\, q_0$ plays the role of an inverse QCD scale, and we take two values of $q_0=0.5$ and 1. These choices are motivated by the need of making the effect of the running of the coupling noticeable in the restricted region of $q$ we explore numerically.

\subsection{Classical solutions}
\label{classical}

The Hamiltonian problem admits a classical solution:
\begin{eqnarray}
\dot{p}=(-\alpha_1+\alpha_2p)p+2(\alpha_3-\alpha_4p) qp\, , \label{eq:2-10}\\
\dot{q}=(\alpha_1-2\alpha_2p)q+(-\alpha_3+2\alpha_4p)q^2\, , \nonumber
\end{eqnarray}
with initial conditions
\begin{equation}
q(y=0)=g_i,\ \ p(y=Y)=g_f\, ,
\label{eq:2-11}
\end{equation}
where $Y$ is  the total rapidity spanned between the projectile located at $y=0$ and the target located at $y=Y$. Thus, the classical amplitude, i.e. the amplitude at tree level, can be computed standardly \cite{Ciafaloni:1978nb,Bondarenko:2006ft}:
\begin{equation}
iA_{fi}^{clas}(y)=1+\sum_k \Delta_k \exp{\left[ -S(Y,q_k,p_k) \right]}\, ,
\label{eq:2-12}
\end{equation}
with the index $k$ running over all possible solutions of the classical equations of motion (\ref{eq:2-10})\footnote{Three solutions, one symmetric ($\Delta_k=+1$) between projectile and target and two asymmetric ($\Delta_k=-1$), appear above some critical rapidity whose value depends on the chosen initial conditions, see a recent discussion in \cite{Bondarenko:2006ft}.},
$\Delta_k=\pm 1$ and $S(Y,q_k,p_k)$ the action evaluated for the classical solutions.

\section{Results}
\label{results}

We begin this section with a description of the numerical method used to compute the transition amplitude.
Following that, we study the rapidity evolution of the solutions for the fan case, RFT, DP and RP. Additionally, we examine the effects of an heuristic introduction of running of the couplings. Further, we phrase the effects of evolution in terms of a saturation scale. Finally, we compare the results of quantum and classical evolution.

\subsection{Numerical method}
\label{numerical}

To solve the differential equation
(\ref{eq:2-2}) with evolution kernel (\ref{eq:2-1}) and initial conditions (\ref{eq:2-3}) which defines the quantum mechanical problem,
we use a second order Runge-Kutta method. We have discretized the $q$-range in 500 points per unit which showed to be enough for a precision better than a few percent. The rapidity region studied is
\begin{equation}
0\leq y \leq 5
\label{eq:3-2}
\end{equation}
(though we study values as high as 40 when comparing with the classical solution).
The step in rapidity is correlated with the
step in $q$ \cite{Braun:2006gy}; we have used
$h=6.25\cdot 10^{-6}$.
With this numerical method we obtain the values of $F_i(y,q)$ in a $q$
interval for different rapidities. The results depends also on the
coupling constants: we use $g_i=1$ \cite{Braun:2006gy}.

The solution to the classical equations of motion (\ref{eq:2-10}) with initial conditions (\ref{eq:2-11}) was obtained using the shooting method. After the classical trajectories were computed, the amplitude was calculated through (\ref{eq:2-12}) (the corresponding quantum expression is given by (\ref{eq:2-4})).

\subsection{Numerical results}
\label{nresults}

We focus on the rapidity evolution of the solutions. 

\renewcommand{\theenumi}{\textbf{(\alph{enumi})}}
\renewcommand{\labelenumi}{\theenumi}
\begin{enumerate}

\item \textbf{fan case}

First, 
as a check, we study the so-called fan case \cite{Schwimmer:1975bv} in which there are no recombination terms, from the projectile point of view, in the Hamiltonian (\ref{eq:2-1}), i.e. 
$\alpha_2=\alpha_4=0$. The differential equation admits an analytical solution
\begin{equation}\label{eq:3-3}
F_{fan}(y,q)=1-\exp\left[-\frac{g_i q e^{\alpha_1 y}}{1+\frac{q\alpha_3}{\alpha_1}\left(e^{\alpha_1 y}-1\right)}\right]\, .
\end{equation}
We use this analytical result to perform a check of our numerical
solution. As we can see in Fig.~\ref{check} there is a very good agreement between the analytical and the numerical results. Note that the BK equation sums the corresponding fan diagrams for BFKL pomerons \cite{Braun:2000wr}, leading a behaviour similar to what we found in this simple zero-dimensional case.

\FIGURE[h]{
      \epsfig{file=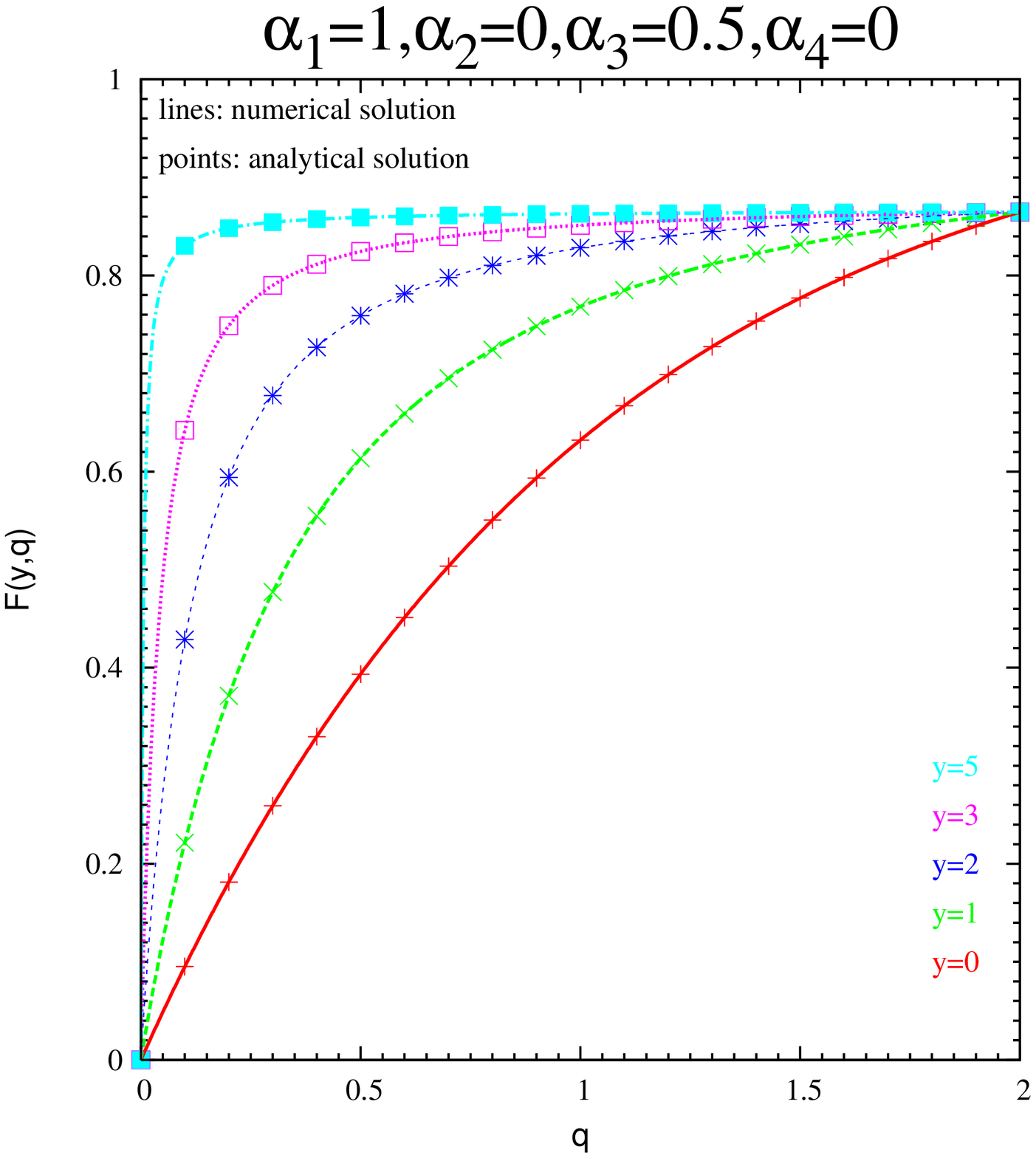,width=9cm}
  \caption{Comparison of the analytical (symbols) and numerical results (lines) for the fan case, with $g_i=1$.}\label{check}
}

\item \textbf{RFT}

Now we turn to RFT -- Fig.~\ref{RFT} -- which contains no quartic vertex. In the limit of high
rapidity we find an exponential decay, a behaviour already found in \cite{Braun:2006gy,Bondarenko:2006rh}
and interpreted as a tunneling phenomenon \cite{Alessandrini:1975ak,Ciafaloni:1977xv,Ciafaloni:1978nb}.

\FIGURE[h]{\epsfig{file=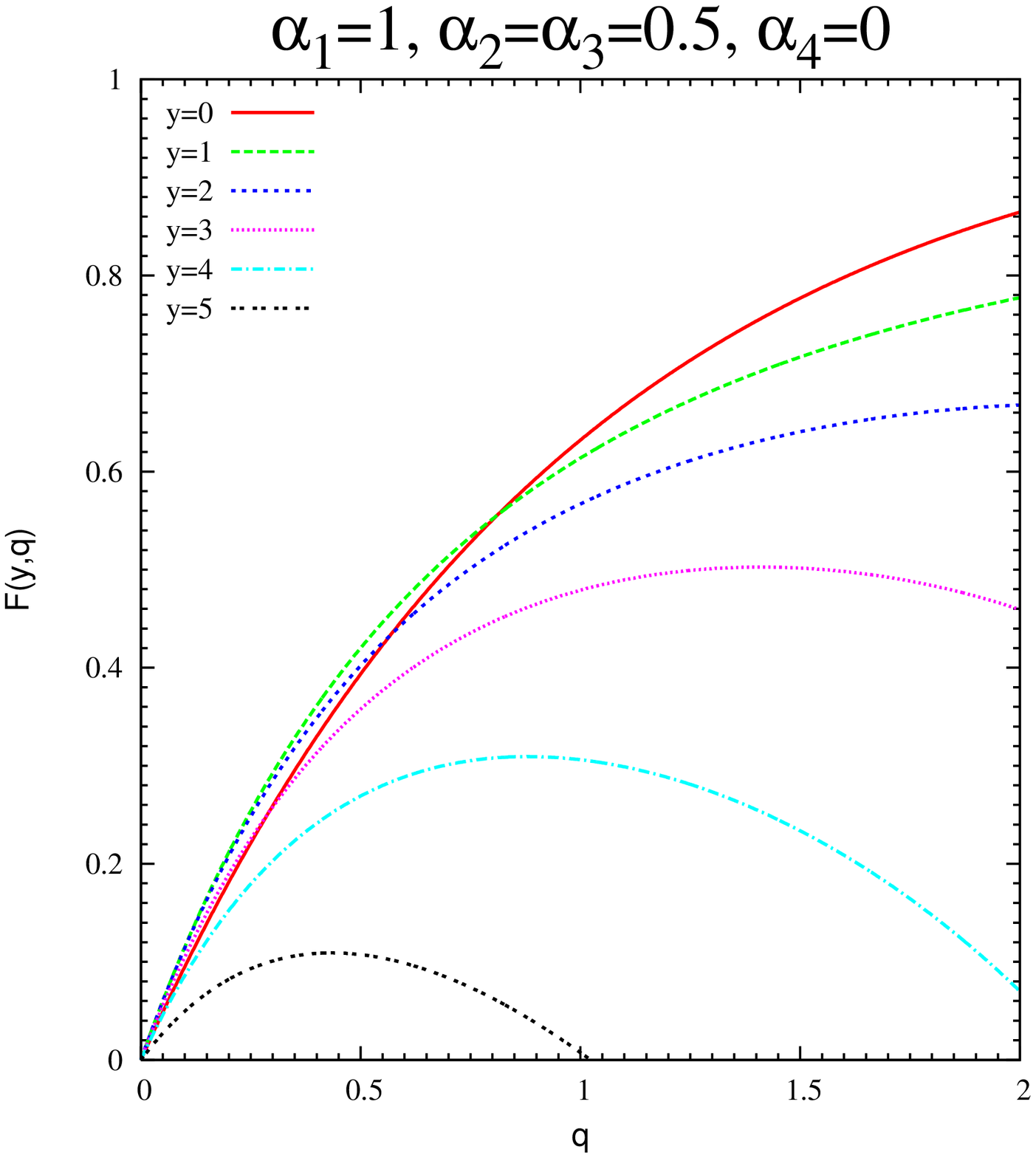,width=9cm}
\caption{Rapidity evolution of the solutions in the case of RFT.}\label{RFT}}

\item \textbf{DP}

The third case we study is DP, with the parameters fixed as discussed in Subsection~\ref{param}.
What we find, Fig.~\ref{DP}, is that the evolution goes in the opposite direction to what is expected in high energy evolution, i.e. that the function moves to smaller values of $q$ with increasing rapidity\,\footnote{Our choice of parameters makes this behaviour visible in the $y$-range we study. Smaller values of $\alpha_{2,3,4}$ make it noticeable only for larger values of $y$.}. This poses serious doubts on DP as a candidate for a description of high energy QCD
evolution. 

\FIGURE[h]{
\includegraphics[width=9cm]{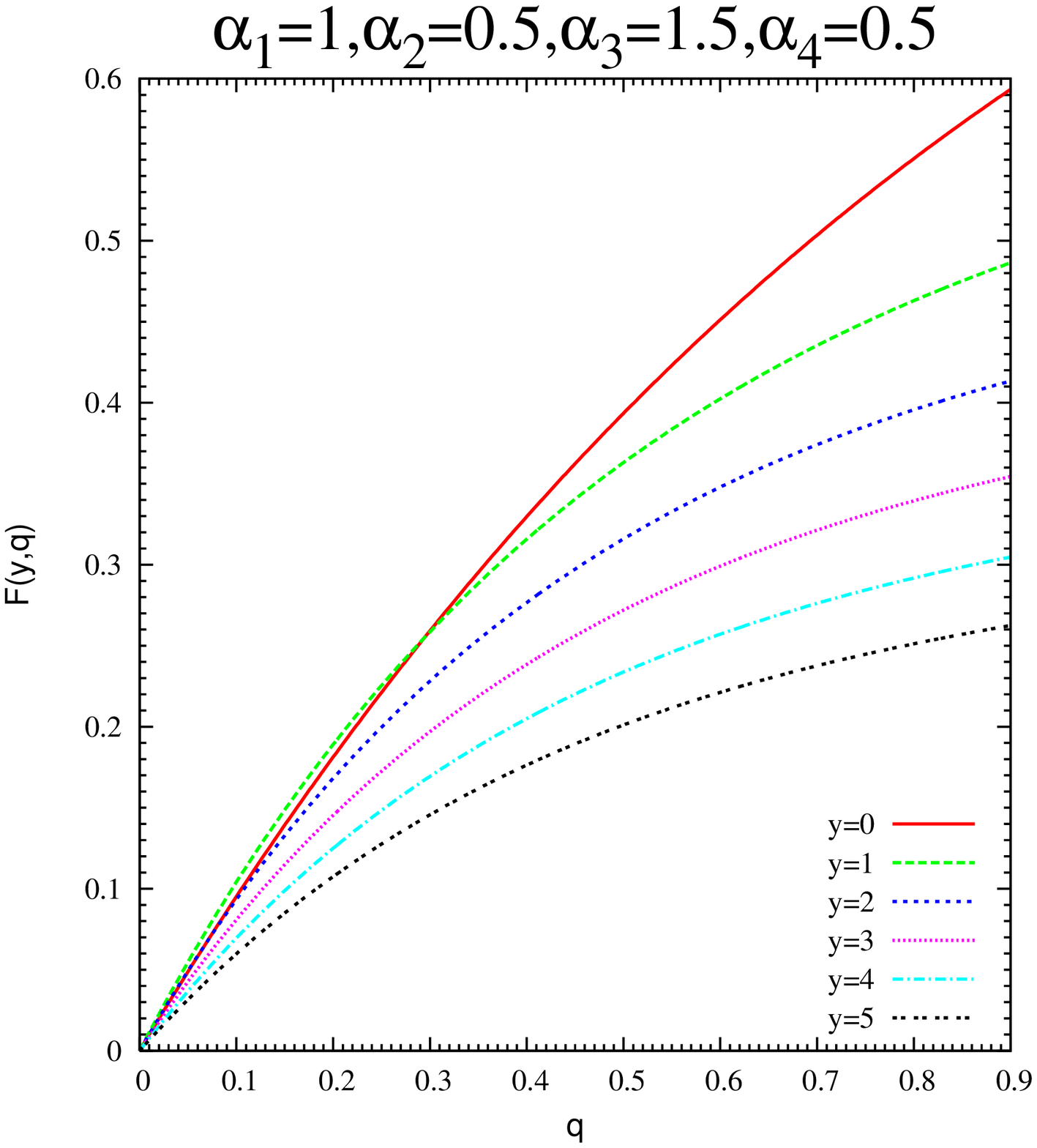}
\caption{Rapidity evolution of the solutions in the case of DP.}\label{DP}
}

\item \textbf{RP}

Finally we turn to RP. As we see in Fig.~\ref{RP} (left),
the evolution leads the front towards smaller values of $q$ as we increase the value of rapidity. 
So it behaves in the expected manner for high energy QCD evolution. 
The parameters $\alpha_{2,4}$, characterizing respectively the vertices $2\rightarrow 1$ and $2\rightarrow 2$, are free.
We find that increasing $\alpha_2=\alpha_4$ makes the evolution softer -- Fig. \ref{RP} (right) --, again as expected for high energy evolution.

\FIGURE[h]{
  \hfill
  \begin{minipage}[t]{7cm}
    \begin{center}
      \includegraphics[width=7cm,clip]{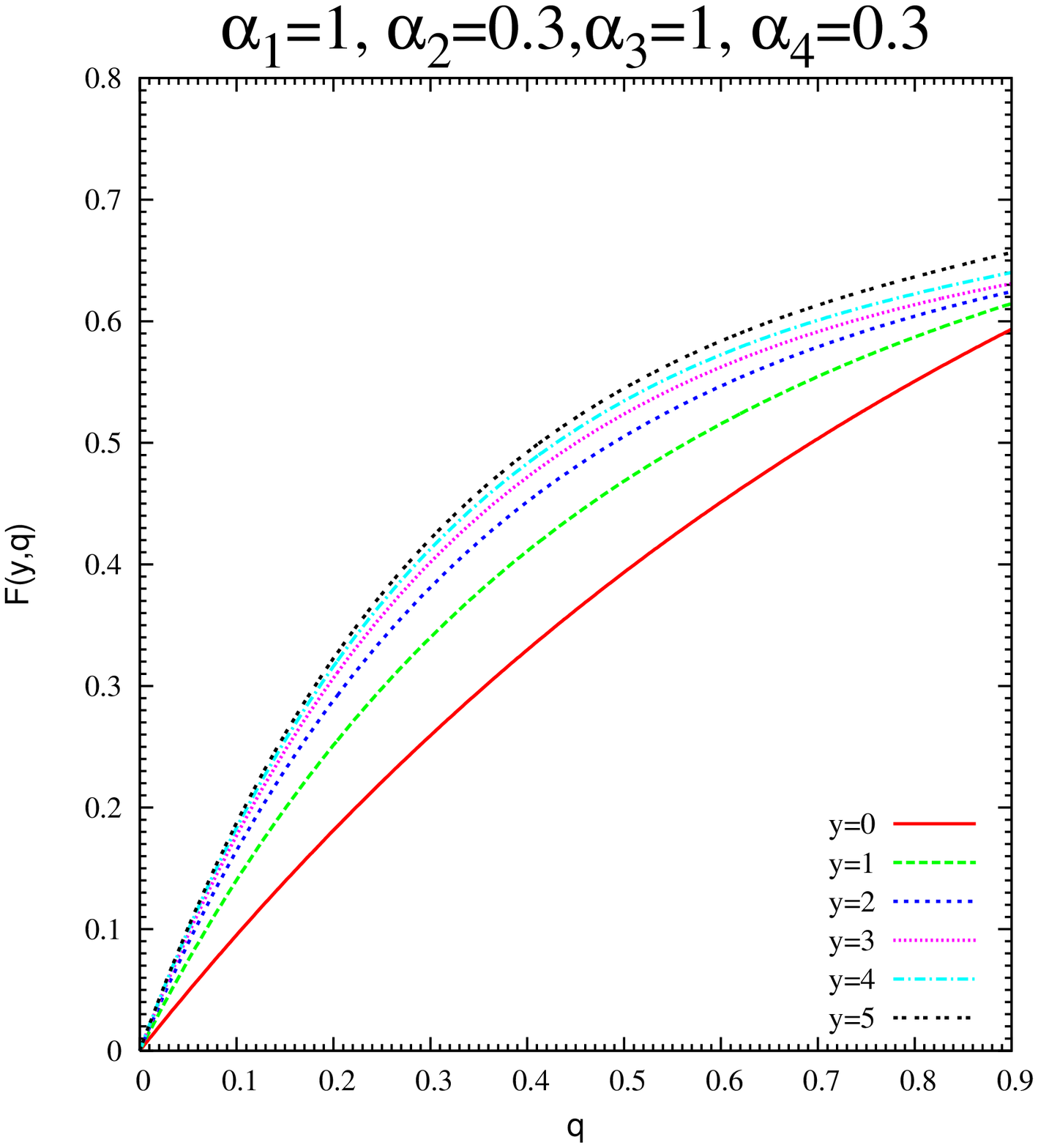}
       \end{center}
  \end{minipage}
  \hfill
  \begin{minipage}[t]{7cm}
    \begin{center}
      \includegraphics[width=7cm,clip]{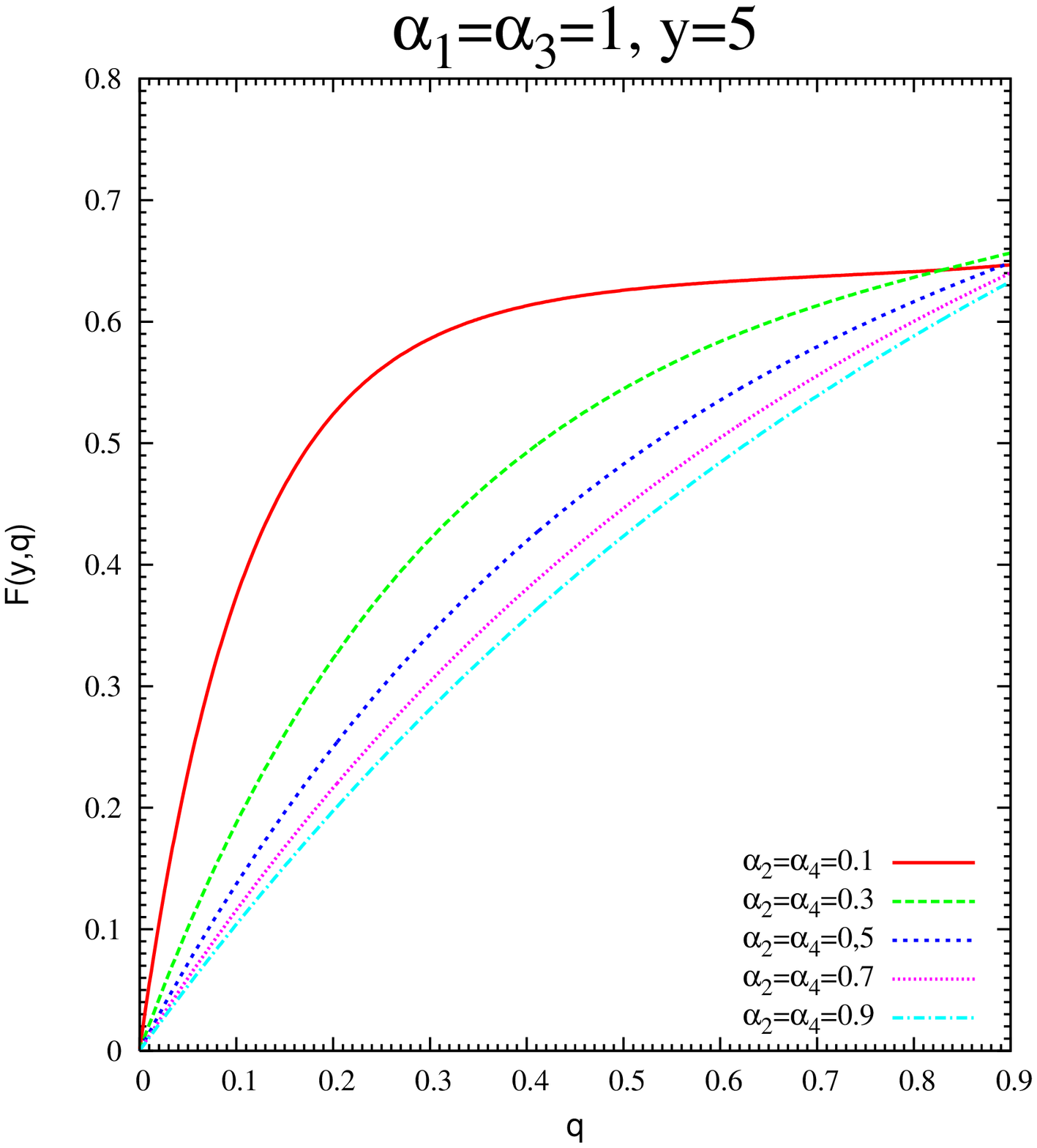}
    \end{center}
  \end{minipage}
  \hfill\caption{Rapidity evolution of the solutions in the case of RP for $\alpha_2=\alpha_4=0.3$ (left). Solutions of RP for different values of $\alpha_2=\alpha_4$ , at $y=5$ (right).}\label{RP}
}

\item \textbf{running coupling}

Now we turn to our heuristic implementation of the running of the couplings. In Fig.~\ref{RPRC} we show the results for the fan case and for RP -- the two cases where the evolution goes in the expected direction -- for the different freezing points for the coupling discussed in subsection \ref{running}. The evolution is clearly slowed down with the effect of the frozen procedure being more noticeable at the beginning of the evolution.

\FIGURE[h]{
  \hfill
  \begin{minipage}[t]{7cm}
    \begin{center}
      \includegraphics[width=7cm,clip]{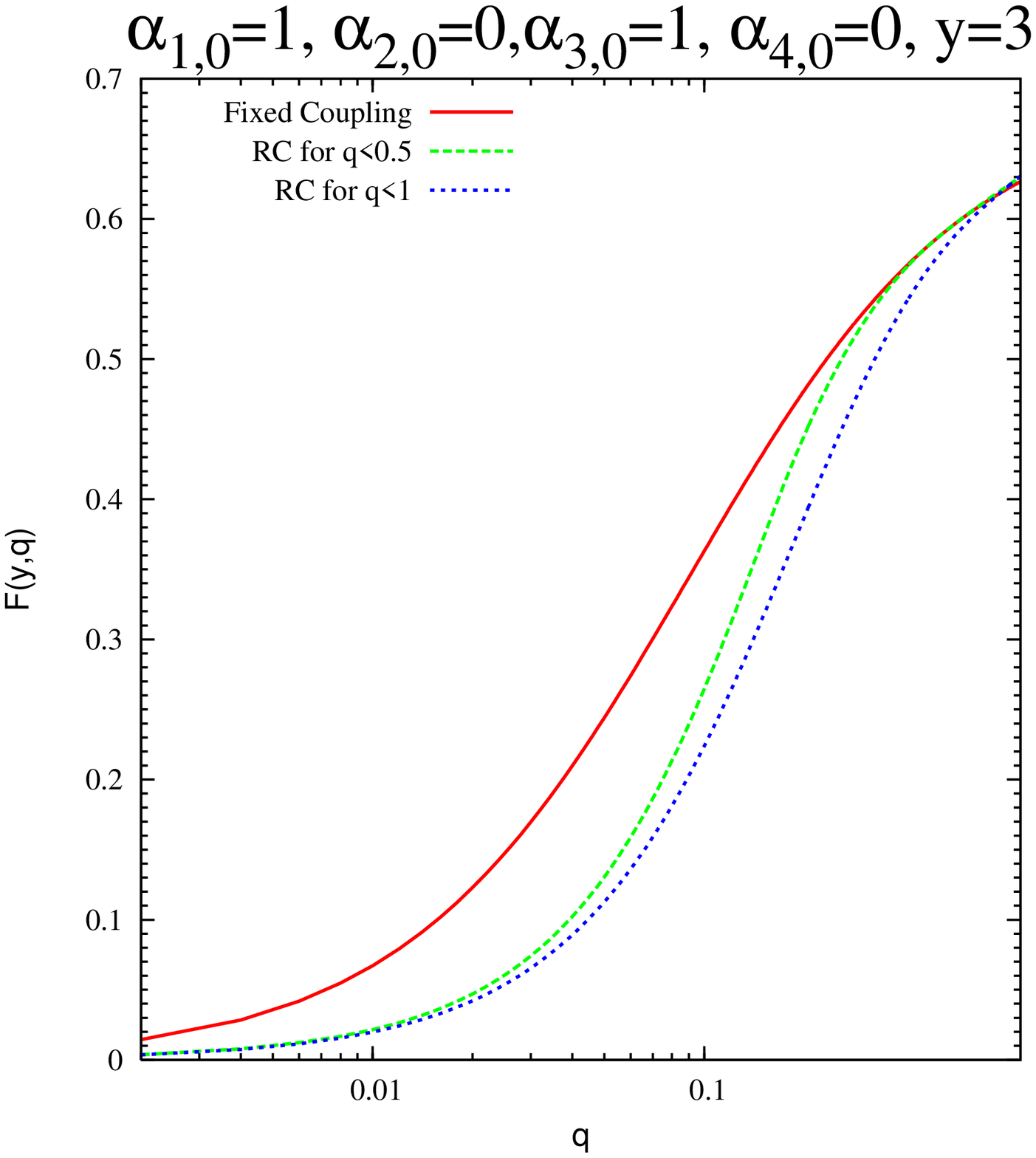}
       \end{center}
  \end{minipage}
  \hfill
  \begin{minipage}[t]{7cm}
    \begin{center}
      \includegraphics[width=7cm,clip]{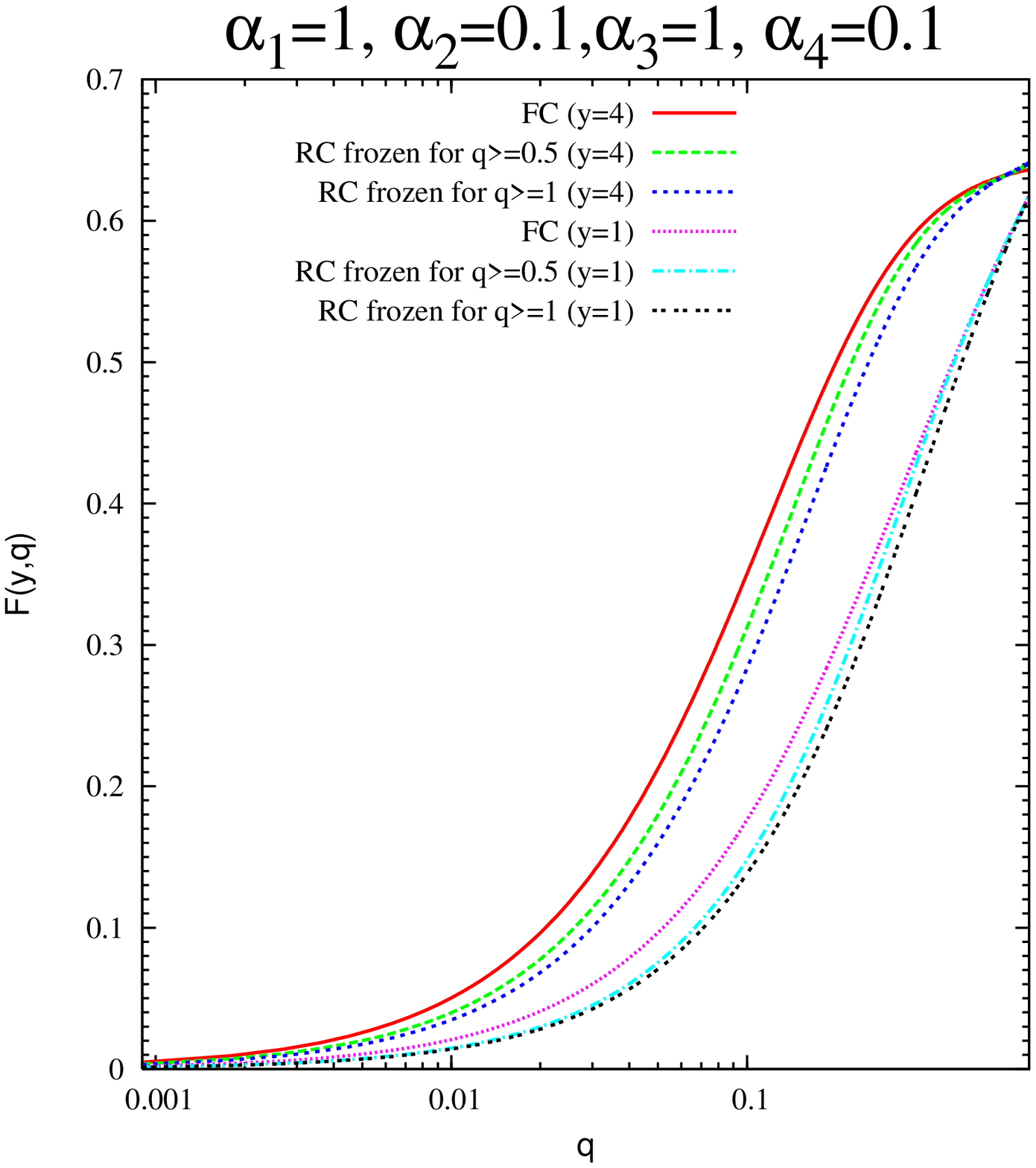}
    \end{center}
  \end{minipage}
  \hfill\caption{Evolution of $F_i(y,q)$ for given values of rapidity,
  comparing the behaviour with fixed (FC) and running (RC) couplings. Fan
  case (left). RP (right).}\label{RPRC}
}

\item \textbf{saturation scale}

To ensure that the effects of evolution in each case are  more easily visible, we parallel the studies of the BK equation and introduce a `saturation scale' $1/q_s(y)$, defined through the equation
\begin{equation}
F(y,q_s(y))=\kappa\, ,
\label{eq:4-1}
\end{equation}
with $\kappa$ some fixed value of order 1/2. 
It is known \cite{Albacete:2004gw} that the specific choice of the value of $\kappa$ leads to small changes in the leading $y$-behaviour of the saturation scale which, for the purposes of this study, are entirely negligible.
In Fig.~\ref{satRC} we show the `saturation scale' obtained for the fan case, for DP and for RP. In the fan case, the running of the couplings slows down the increase of $1/q_s(y)$ (as in the BK equation). In DP $1/q_s(y)$ diminishes with increasing rapidity. Finally, in RP both the increase of the strengths of the vertices $2\rightarrow 1$ and $2\rightarrow 2$ and the running of the couplings slow down the increase of  $1/q_s(y)$ -- a behaviour which may be linked with the findings in \cite{Dumitru:2007ew} of a competition between the effects of recombination vertices, called `Pomeron loops', and of running coupling.



\FIGURE{
      \includegraphics[width=7cm,clip]{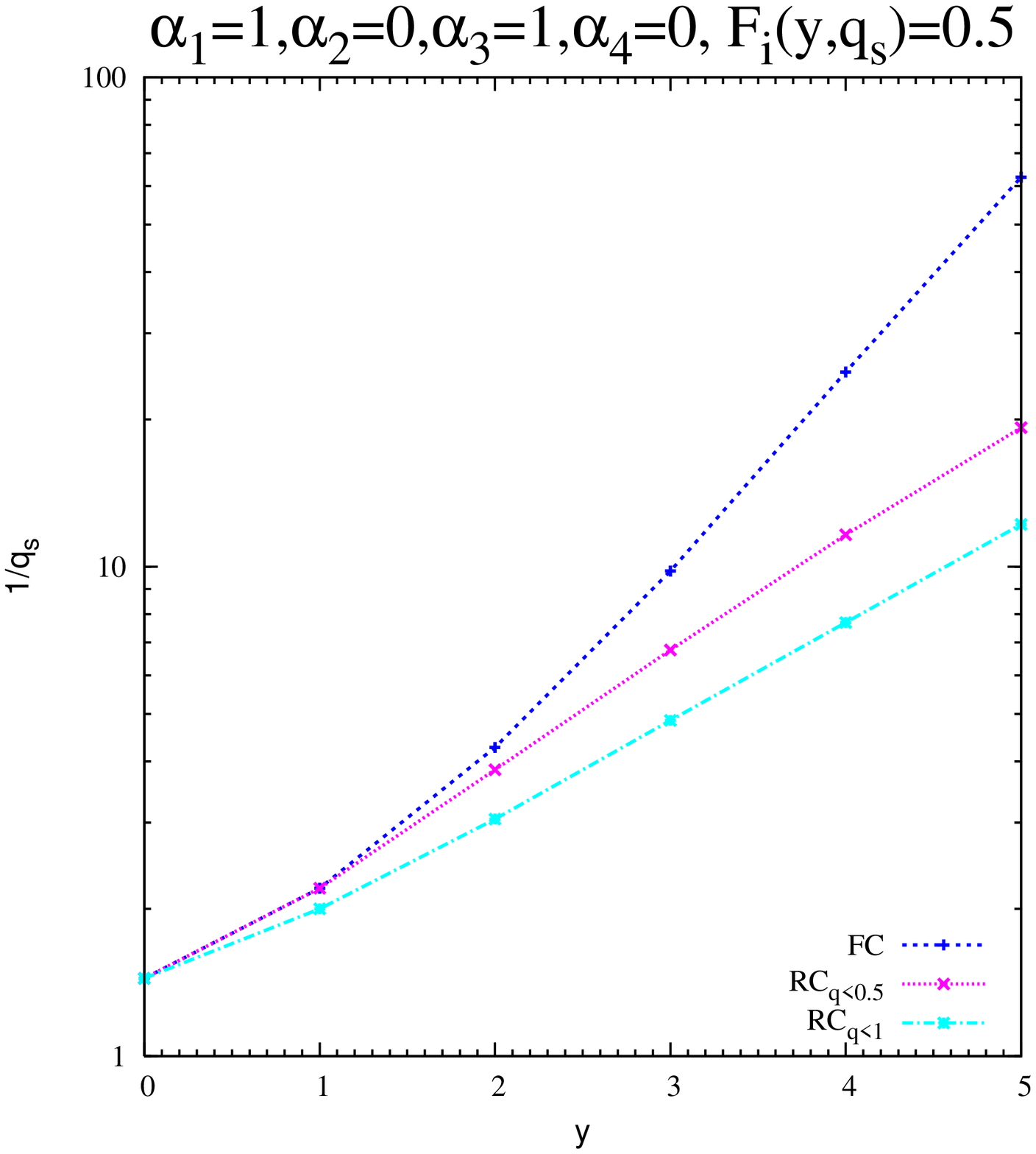}
       \includegraphics[width=7cm,clip]{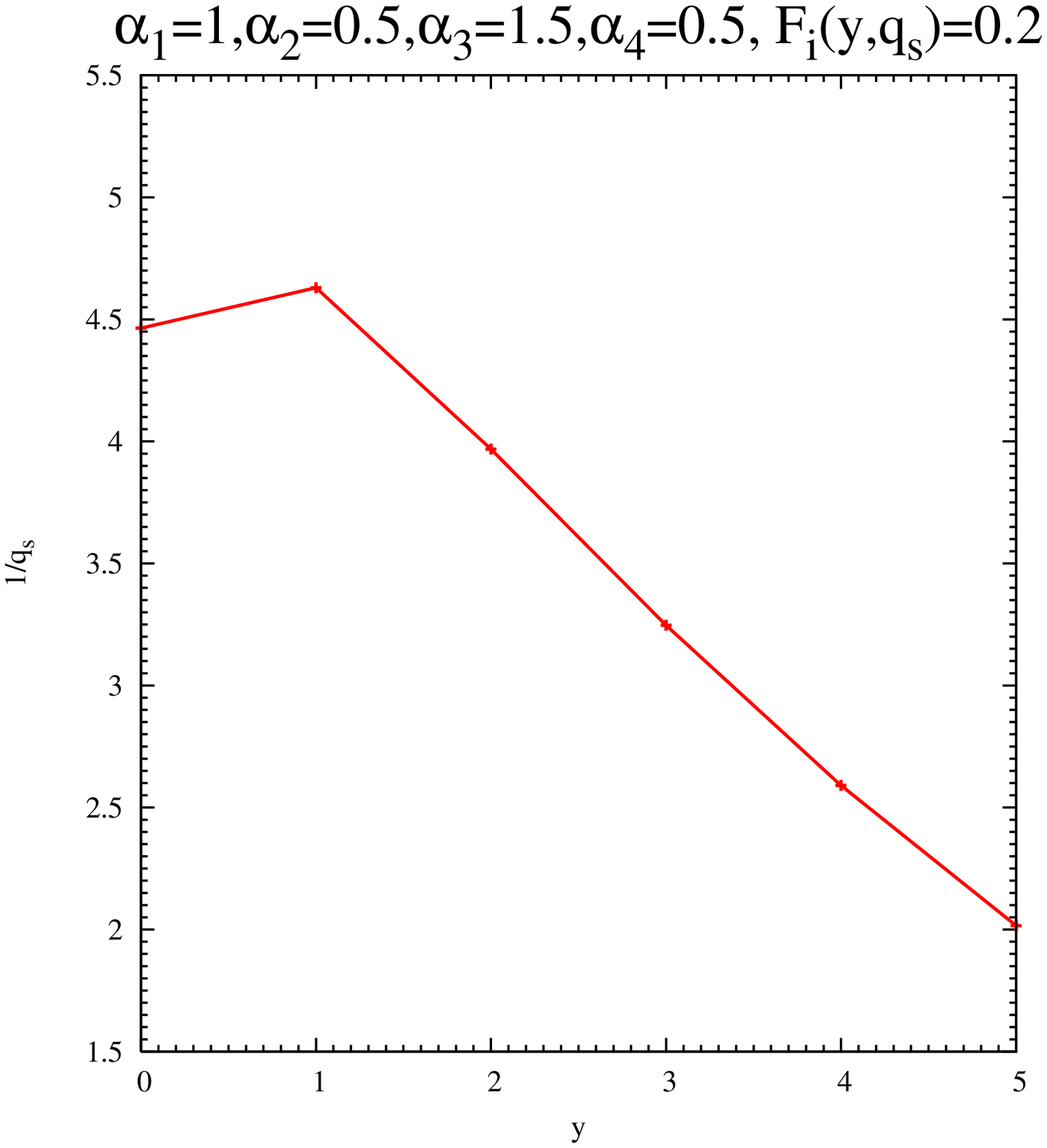}
      \includegraphics[width=7cm,clip]{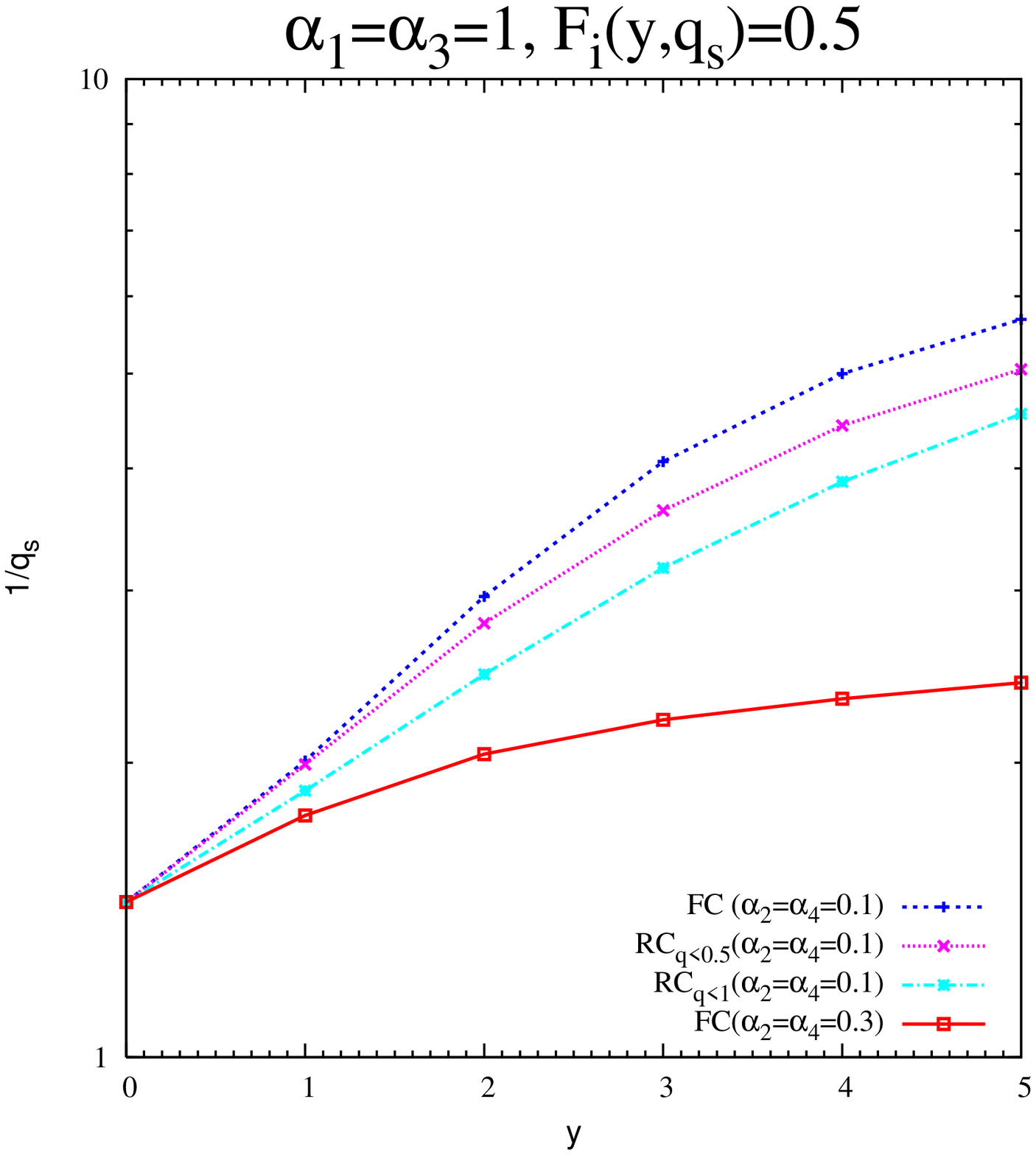}
  \caption{$1/q_s(y)$ in the fan case for fixed and running couplings (top left). $1/q_s(y)$ in DP for fixed coupling (top right). $1/q_s(y)$ in RP with fixed coupling for two sets of $\alpha_2=\alpha_4=0.1$ and 0.3, and for running coupling for $\alpha_2=\alpha_4=0.1$ (bottom).}\label{satRC}
}

\item \textbf{classical solution}

Finally, we want to compare the full quantum and the symmetrical classical solutions, see Subsection \ref{classical}. This provides an evaluation of the effect of loops in the strict Quantum Field Theory sense.
In Fig.~\ref{class0.1} we show both the classical and quantum imaginary parts of the amplitude for RFT with a quartic vertex, for different values of $g_i=g_f$. The classical solution lies always above the quantum one, with the difference between both increasing with increasing rapidity. Similar conclusions are extracted in other cases.

\FIGURE{
      \includegraphics[width=7cm,clip]{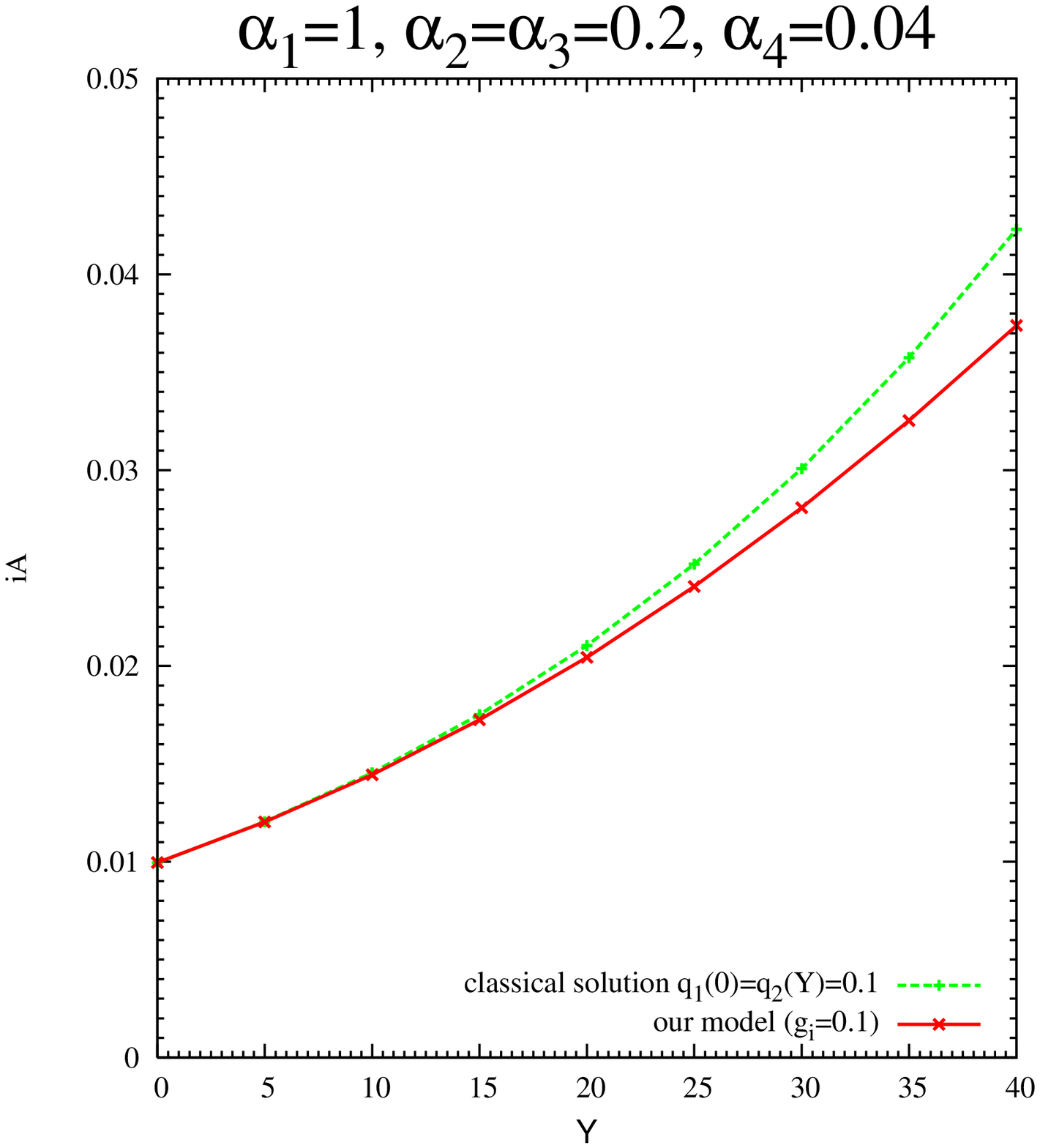}
      \includegraphics[width=7cm,clip]{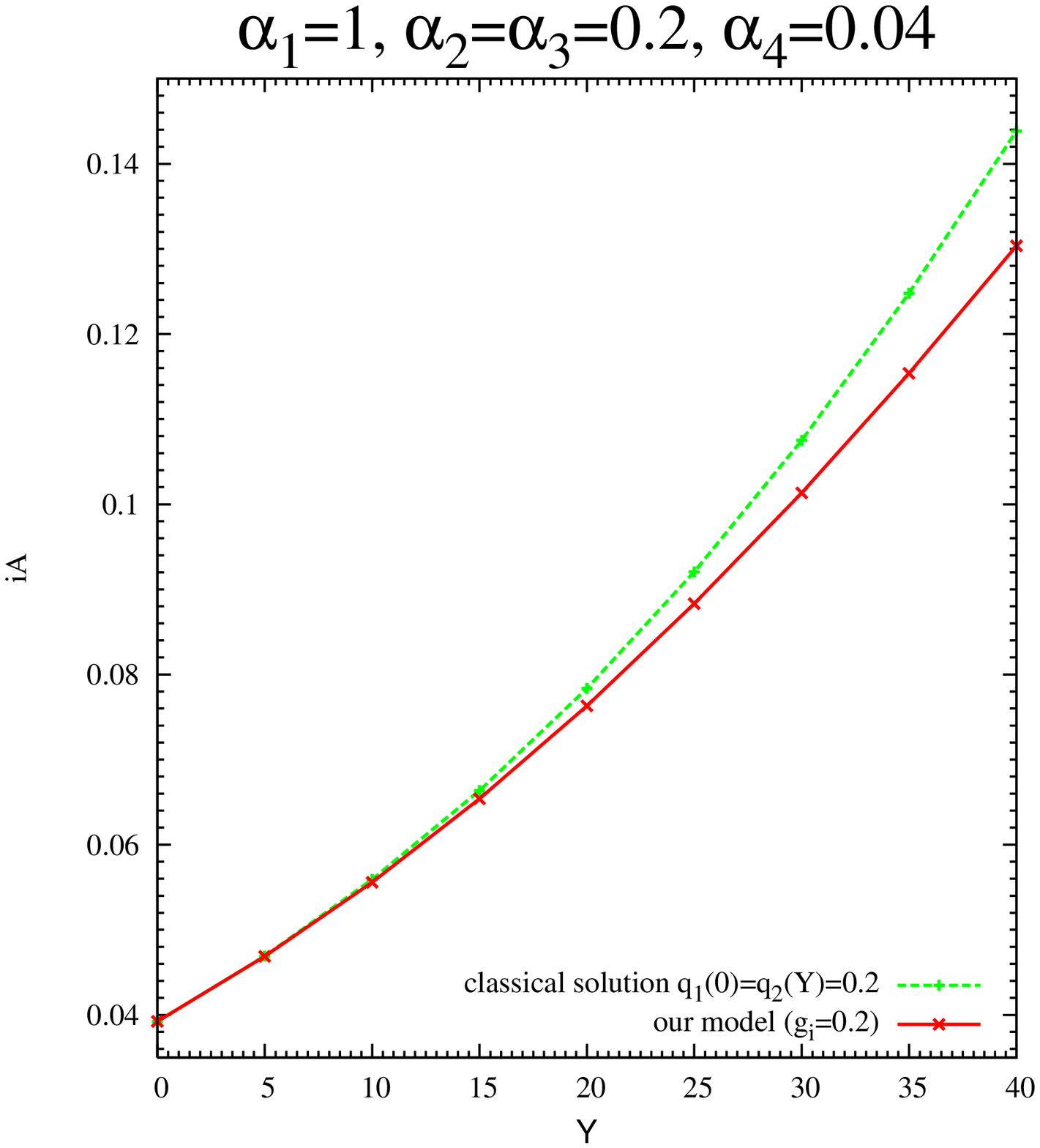}
      \includegraphics[width=7cm,clip]{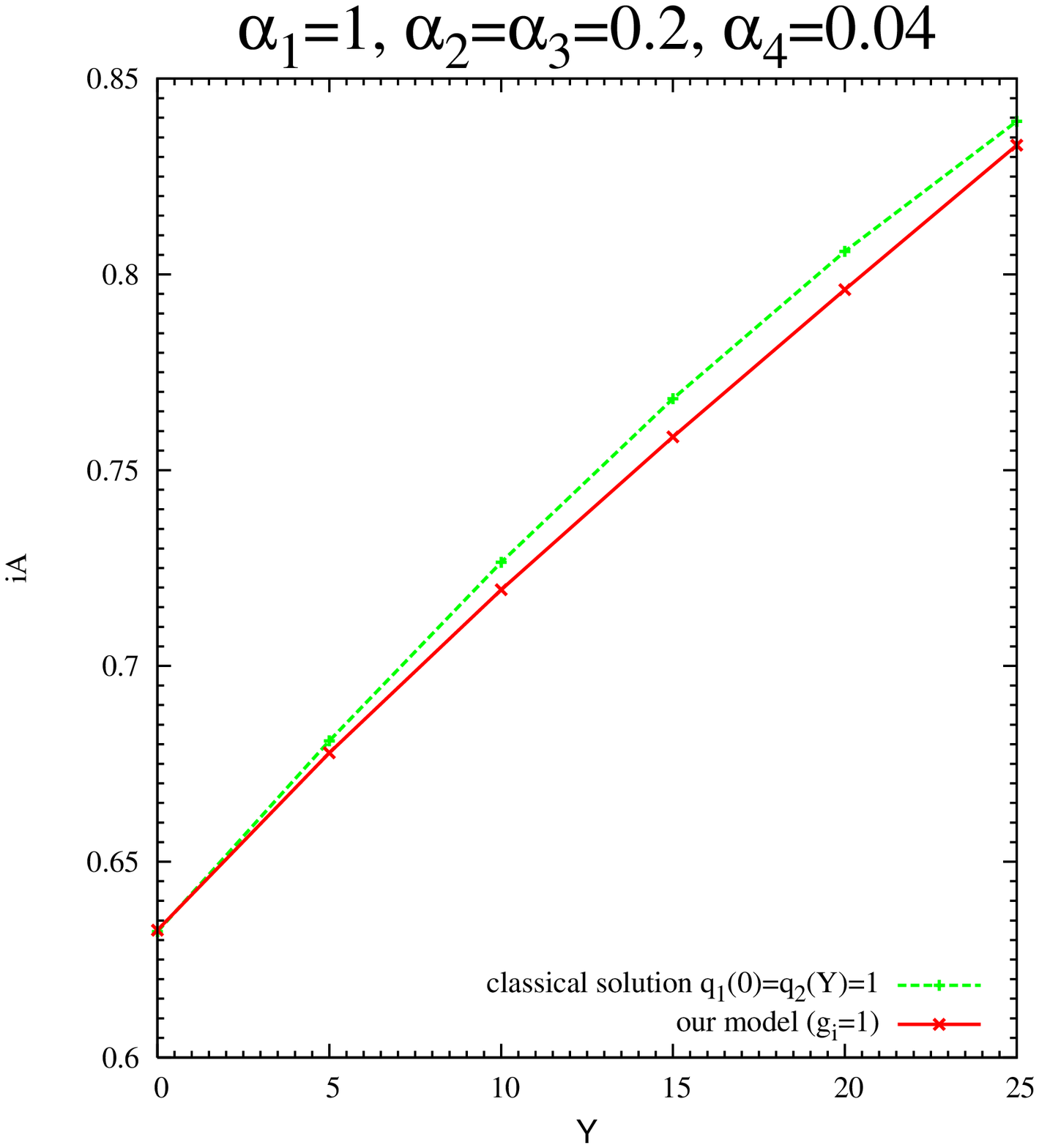}
   \caption{The classical symmetrical (dashed lines) and the quantum (solid line) imaginary part of the amplitude versus $y$. $g_i=g_f=0.1$ (top left).  $g_i=g_f=0.2$ (top right). $g_i=g_f=1$  (bottom).}\label{class0.1}
}

%

%

\end{enumerate}

\section{Conclusions}
\label{conclusions}

In this paper we have examined numerically several reaction-diffusion processes in zero transverse dimensions as possible candidates for a toy model of high-energy QCD evolution. We have restricted ourselves to those models in which the Hamiltonian formulation \cite{elka} contain vertices with up to two derivatives: diffusion, splitting, recombination and $2\rightarrow 2$, which allows a discussion of all common universality classes. We have also considered usual variants of such models which are not strictly reaction-diffusion processes.
Their feasibility as toy models for high-energy QCD evolution has been evaluated in terms of the direction of the evolution with increasing evolution parameter, that is increasing rapidity.

In the case of Reggeon Field Theory, which contains no quartic vertex and thus has no reaction-diffusion interpretation, we find the known behaviour of a vanishing amplitude with increasing rapidity. Directed Percolation leads to a behaviour of the solutions decreasing with increasing rapidity. On the other hand, the limiting case of considering only fan diagrams (again with no reaction-diffusion counterpart) and Reversible Processes show solutions which increase with increasing rapidity. In the case of RP which, as DP, contains a quartic vertex, the evolution is slowed down by increasing the recombination and $2\rightarrow 2$ vertices. We have shown all these behaviours both at the level of the amplitudes and of an analogue of a saturation momentum.

We have also introduced, in an heuristic manner, a running of the couplings, which tends to slow down the evolution in all cases, working in the same direction as the increase of the recombination and $2\rightarrow 2$ vertices as found in \cite{Dumitru:2007ew}. Finally, we have numerically computed the classical solutions (those which sum tree diagrams), and found that they generically lie above the quantum ones, with their difference growing with increasing rapidity, so quantum effects tend to slow down the evolution.

Summarizing, we have found that the only zero-dimensional reaction-diffusion process which shows a behaviour of the amplitude in agreement with the expectations from high energy QCD with rapidity is RP. This process is linked with sFKPP equation often used to discuss \cite{Iancu:2004iy,Iancu:2005nj,Mueller:2005ut,Bondarenko:2007kg} high-energy QCD evolution. Increasing recombinations terms, and the inclusion of quantum loops and of a running of the coupling generically slow down the evolution.

\section*{Acknowledgments}
It is a pleasure to thank E. Iancu, A. Kovner, M. Kozlov, E. M. Levin, A. H. Mueller and A. Prygarin for most useful discussions. Special thanks go to M. A. Braun for discussions and a numerical cross-check, to L. Motyka and  R. Peschanski for most useful remarks. NA has been supported by Ministerio de Educaci\'on
y Ciencia of Spain under a contract Ram\'on y Cajal, and PQ by MEC of Spain under a grant of the FPU Program. NA, SB and PQ have been supported by MEC of Spain under project FPA2005-01963, by Xunta de Galicia (Conseller\'{\i}a de Educaci\'on), and by the Spanish Consolider-Ingenio 2010
Programme CPAN (CSD2007-00042). JGM has been supported by
the Funda\c c\~ao para
a Ci\^encia e a Tecnologia of Portugal under contracts
SFRH/BPD/12112/2003 and SFRH/BPD/34760/2007. JGM thanks the Departamento de 
F\'{\i}sica de Part\'{\i}culas of Universidade de Santiago de Compostela, and
NA and PQ thank CENTRA/IST, for warm hospitality during stays
when part of this work was done.


\providecommand{\href}[2]{#2}\begingroup\raggedright\endgroup

\end{document}